\documentclass[aps,prl,twocolumn,superscriptaddress,showpacs,draft]{revtex4}
\usepackage{graphicx}
\input{epsf}

\begin{document}

\title{Chaos at Fifty Four in 2013}

\author{D.L.Shepelyansky}
\affiliation{\mbox{Laboratoire de Physique Th\'eorique du CNRS, IRSAMC, 
Universit\'e de Toulouse, UPS, F-31062 Toulouse, France}}

\date{ Jun 27, 2013}


\begin{abstract}
In contrast to the claim of Motter and Campbell on Chaos at Fifty,
done at arXiv:1306.5777, it is pointed out that in 2013 we  are at Chaos Fifty Four,
if to count correctly and to remember about pioneering results of 
Boris Chirikov obtained in 1959.
\end{abstract}

\maketitle

The recent Feature Article entitled {\it Chaos at Fifty} by
Motter and Campbell \cite{campbell}
highlights Lorenz's discovery in 1963, which,
as they say, ``gave birth to a field that still thrives''.
Without any doubts Edward Lorentz did 
an outstanding work but the real history of scientific research
of chaos does not allow to attribute the birth event to
Fifty in 2013.

\begin{figure}
\centerline{ \epsfxsize=4.0cm\epsffile{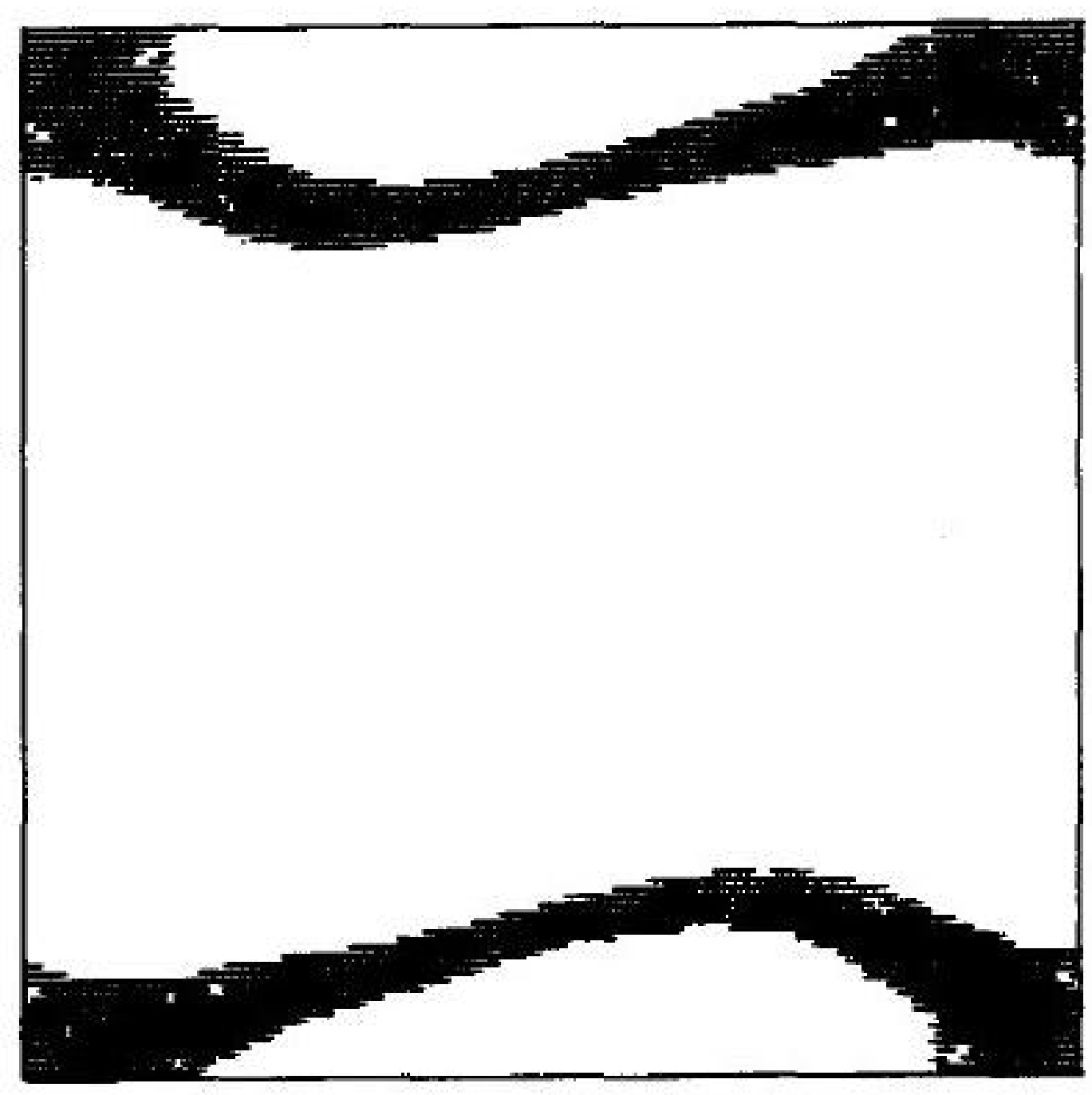}
\hskip 0.1cm  \epsfxsize=4.0cm\epsffile{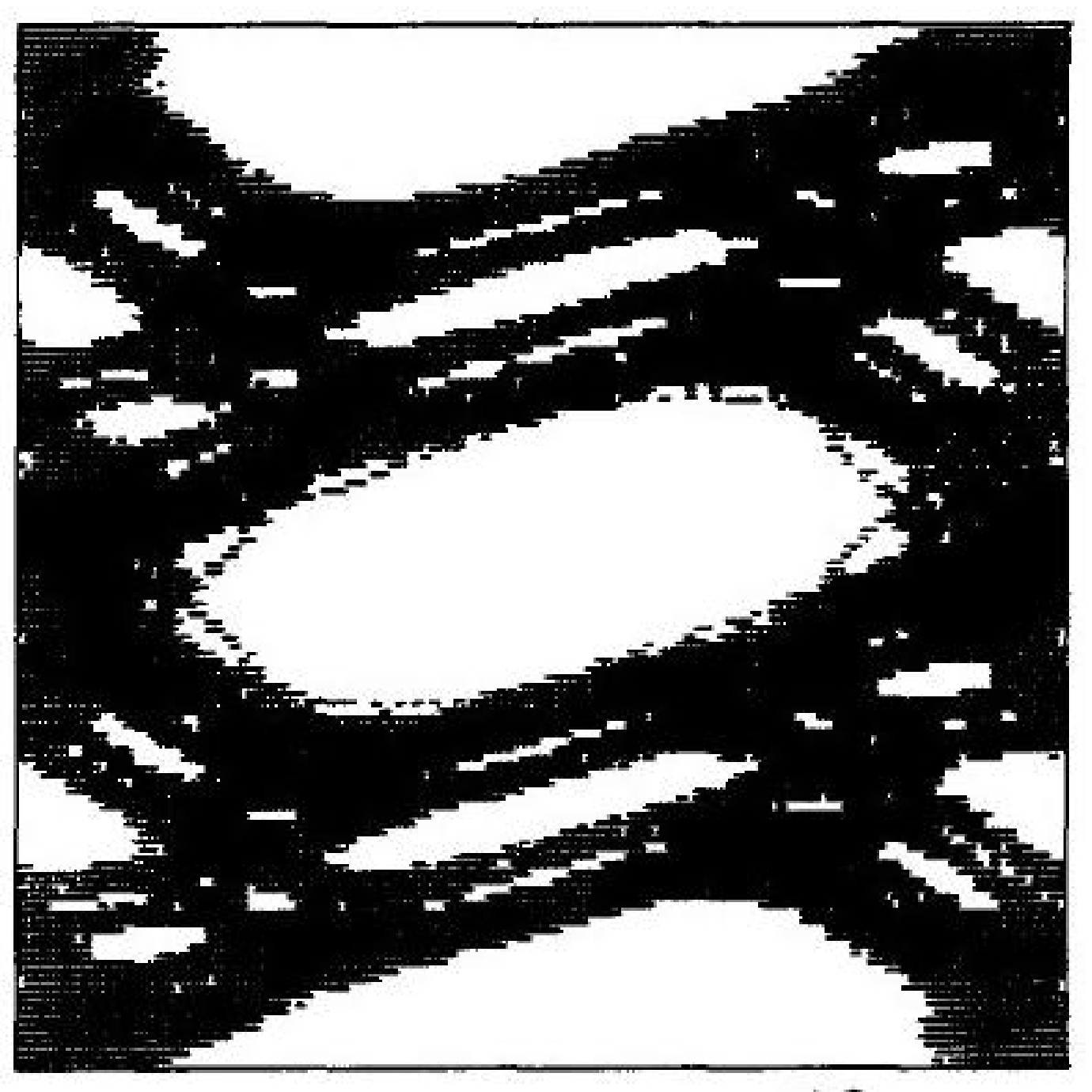}
}
\vglue -0.2cm
\caption{Poincar\'e section of the phase space $(x,y)$ of the
Chirikov standard map at chaos parameter
$K=0.96$ (left), $K=1.13$ (right);
chaos regions covered by one trajectory
are shown in black (after \cite{chirikov1969,chirikov1979}).
The map has the form $\bar{y}=y+K \sin x, \bar{x}=x+\bar{y} (mod 2\pi)$,
where bars mark new variable values after one map iteration.
} 
\label{fig1}
\end{figure}

The physical birth of chaos happened before that in Hamiltonian systems
with a few degrees of freedom.
In fact, the Chirikov resonance-overlap criterion \cite{chirikov1959}
was introduced in 1959 by Boris Chirikov and successfully 
applied by him to explain the confinement border for plasma 
in open mirror traps observed in experiments at the Kurchatov Institute
at Moscow. This was the very first physical and 
analytical criterion for the onset of chaotic motion 
in deterministic Hamiltonian systems. 
Its validity in generic Hamiltonian systems has been
confirmed in numerous numerical simulations
performed by Chirikov and his collaborators
for the Fermi acceleration model
\cite{chirikov1964}, the Fermi-Pasta-Ulam problem \cite{chirikov1966},
dynamical maps and various physical systems 
\cite{chirikov1969,chirikovufn},
\cite{chirikov1979}.
An example of resonance overlap 
with integrable islands and chaotic component of
motion is shown in Fig.~\ref{fig1}
for the Chirikov standard map.
These figures are take from the review papers of Chirikov
\cite{chirikov1969,chirikov1979}. 
The phase space of the same map 
at similar parameters is shown
in Fig.~4 of \cite{campbell}
with computer facilities
being much more advanced 
compared to those of 1969 and 1979.
Unfortunately, the authors of \cite{campbell}
forget to say that the main properties and universal
features of this map had been 
discovered  by Chirikov
(read more about the Chirikov standard map at \cite{stmap}).
This map describes the confinement border for plasma 
in open mirror traps explained in \cite{chirikov1959}.

The recognition of pioneering results of 
Chirikov is well established among experts of chaos community. 
In contrast to \cite{campbell},
this is the most quoted author in the fundamental 
book on regular and chaotic dynamics \cite{lichtenberg}.
It should be also pointed out that
there are also many fundamental mathematical and physical
papers on dynamical chaos, done in USSR-Russia and Western Europe, 
which remained  ignored in \cite{campbell}.
These works, done at the early stage of chaos research,
include the Kolmodorov-Arnold-Moser theory
of integrability \cite{kam1,kam2},\cite{kam3}, chaos in
Anosov systems \cite{anosov} and Sinai billiards \cite{sinai},
Henon-Heiles model \cite{henon},
Kolmogorov-Sinai entropy \cite{ks} and others.

A more balanced description of history of dynamical chaos
can be find in \cite{lichtenberg,ks,holmes}.
Articles about scientific research of Boris Chirikov
are available at \cite{bellissard,physicad},
\cite{phystoday,bvchirikov},\cite{criterion}
and at the website dedicated to him \cite{webpage}.

\end{document}